\documentclass[12pt]{spieman}  
\usepackage{amsmath,amsfonts,amssymb}
\usepackage[pdftex]{graphicx}
\usepackage{setspace}
\usepackage{tocloft}

\usepackage{textcomp}
\usepackage{float}
\usepackage{xspace}
\usepackage{threeparttable}
\usepackage{multirow}
\usepackage{url}
\usepackage{arydshln}

\newcommand{\um}{\textmu m\xspace}
\newcommand{\ug}{\textmu g\xspace}

\newcommand{\rmalpha}{\rm $\alpha$}

\title{X-ray transmission calibration of the gate valve for the X-ray astronomy satellite XRISM}

\author[a,b*]{Takuya Midooka}
\author[a]{Masahiro Tsujimoto}
\author[c]{Shunji Kitamoto}
\author[a,d]{Nozomi Nakaniwa}
\author[a]{Yoshitomo Maeda}
\author[a,d]{Manabu Ishida}
\author[a,b]{Ken Ebisawa}
\author[a,b]{Mayu Tominaga}
\affil[a]{Japan Aerospace Exploration Agency (JAXA), Institute of Space and Astronautical Science (ISAS), 3-1-1 Yoshinodai, Chuo, Sagamihara, Kanagawa, Japan, 252-5210}
\affil[b]{The University of Tokyo, 7-3-1 Hongo, Bunkyo, Tokyo, Japan, 113-8654}
\affil[c]{Rikkyo University, 3-3-4-1 Nishi-ikebukuro, Toyoshima, Tokyo, Japan, 171-8501}
\affil[d]{Tokyo Metropolitan University, 1-1 Minami-Osawa, Hachioji, Tokyo, Japan, 192-0397}

\cftpagenumbersoff{figure}
\cftpagenumbersoff{table} 
\begin{document} 
\maketitle

\begin{abstract} 
\textit{Resolve} onboard the X-ray satellite XRISM is a cryogenic instrument with an
 X-ray microcalorimeter in a Dewar. A lid partially transparent to X-rays (called gate
 valve, or GV) is installed at the top of the Dewar along the optical axis. Because
 observations will be made through the GV for the first few months, the X-ray
 transmission calibration of the GV is crucial for initial scientific outcomes. 
 We present the results of our ground calibration campaign of the GV, which is composed
 of a Be window and a stainless steel mesh.  For the stainless steel mesh, we measured
 its transmission using the X-ray beamline at ISAS.  For the Be window, we used
 synchrotron facilities to measure the transmission and modeled the data with (i)
 photoelectric absorption and incoherent scattering of Be, (ii) photoelectric absorption
 of contaminants, and (iii) coherent scattering of Be changing at specific energies.  We
 discuss the physical interpretation of the transmission discontinuity caused by the
 Bragg diffraction in poly-crystal Be, which we incorporated into our transmission
 phenomenological model. We present the X-ray diffraction measurement on the sample to
 support our interpretation. The measurements and the constructed model meet the
 calibration requirements of the GV.
 We also performed a spectral fitting of the Crab nebula observed with Hitomi SXS and
 confirmed improvements of the model parameters.
\end{abstract} 

\keywords{XRISM, \textit{Resolve}, Beryllium, gate valve, X-ray transmission, Bragg diffraction}

{\noindent \footnotesize\textbf{*}Takuya Midooka,  \linkable{midooka@ac.jaxa.jp} }

\begin{spacing}{2}   

\section{Introduction}
\label{sec:intro} 
The X-ray astronomy satellite XRISM is planned to be launched in the Japan fiscal year
2022\cite{Tashiro20}. It is expected to revolutionize high-energy astrophysics with its
high-resolution, high-throughput, and non-dispersive X-ray spectrometer based on the
X-ray microcalorimetry. The instrument called \textit{Resolve} is built on the heritage
of Hitomi SXS\cite{Kelley16}, which demonstrated its capability in orbit but was
discontinued a month after launch due to unexpected failure of the spacecraft attitude
control system in 2016 \cite{Takahashi18}.

\textit{Resolve} is a cryogenic instrument with an X-ray microcalorimeter detector in a
Dewar \cite{Ishisaki18, Fujimoto17}. To maintain the Dewar vacuum on ground and to
protect the detector from the initial spacecraft outgassing in orbit, a partially
transparent lid called the gate valve (GV) is installed at the top of the Dewar along
the X-ray optical path.  GV is composed of a Be window and a stainless steel mesh (see
details in Section~\ref{sec:gv}).  On ground, the GV can be opened and closed
repeatedly. In orbit, it can be opened only once using a non-explosive actuator system
and cannot be closed once opened. It is planned to be opened during the commissioning
phase when the outgassing settles. Until then, all observations will be made through the
GV. Therefore, calibrating the energy dependence of the X-ray transmission of this
apparatus is crucial for the initial scientific outcomes.

Be windows are routinely used as a standard entrance window in X-ray detectors requiring
a leak-tight structure. In the past, the modeling based only on the Be photo-electric
absorption was sufficient (e.g. ASCA GIS\cite{Ohashi96}). For X-ray spectrometers with a
high spectral resolution and a high throughput such as Hitomi SXS\cite{Kelley16}, XRISM
\textit{Resolve}\cite{Ishisaki18}, and Athena X-IFU\cite{Barret18}, this is
insufficient. This was recognized in SXS \cite{Yoshida17}, based on which we develop the
model for \textit{Resolve} in this work.

In the SXS, the GV was not calibrated using the flight model on ground unlike the other
components \cite{Eckart18}. This choice was motivated by the lower priority due to its
not being used under nominal observations and some scheduling constraints. Due to the
unexpected early loss of the spacecraft, all observations were made with the GV
closed. After the loss, the transmission measurements were performed using a spare unit
\cite{Eckart18, Yoshida17, Hoshino17}. As a result, the GV turned out to be the cause
for the largest systematic uncertainty in the effective area calibration in the Crab
observation with the SXS \cite{Tsujimoto18}.

Based on this lesson, we planned to perform the ground calibration measurements for all
parts of the GV for \textit{Resolve} with a clear set of calibration requirements that
are traced back to science goals of the mission.  We aim to understand causes of the
large systematic uncertainty in the SXS GV and reduce it.  The allocated budgets for the
GV are shown in Table~\ref{tab:requirements}.  The calibration results will be delivered
to users through the calibration database (CALDB).

\begingroup
\renewcommand{\arraystretch}{1.3}
\begin{table}[H]
\caption{Calibration requirements of the Be window and the stainless steel mesh\cite{Eckart19req}}\label{tab:requirements}
\scalebox{0.72}{
\begin{tabular}{cccccl}\hline
Component & Items & Accuracy & Range (req) & Range (goal) &\multicolumn{1}{c}{Comment}\\\hline\hline
Be window& \begin{tabular}{c}Broadband\\\relax (absolute transmission) \end{tabular} & $\pm$1\% & 2.5--12.0~keV & 1.8--25.0~keV & \\\cdashline{2-6}[3pt/1.4pt]
& \begin{tabular}{c}Fine structures\\\relax (relative transmission) \end{tabular} & $\pm$5\% & & & \begin{tabular}{l}BDFs and contaminants\\ Knowledge of structure to 2~eV required\end{tabular}\\\cdashline{2-6}[3pt/1.4pt]
& \begin{tabular}{c}Spatial uniformity\\\relax (relative transmission) \end{tabular} & $\pm$3\% & \begin{tabular}{c}single energy\\ ($\sim$3.6~keV)\end{tabular} && \begin{tabular}{l}With $>$100 spatial grids\\ \end{tabular}\\\hline
Mesh & Broadband & $\pm$1\% & 2.5--12.0~keV & 1.8--25.0~keV & \\
\hline
\end{tabular}
}
\end{table}
\endgroup

We performed a ground calibration campaign using various facilities in 2018--2019 and
achieved the required calibration accuracy before the GV was installed in the flight
model (FM) of \textit{Resolve} at the end of 2019. In this paper, we describe the
details of this campaign in the following structure. First, we give a brief description
of the GV in Section~\ref{sec:gv}. We then present the results of the measurements in
Section~\ref{sec:results}. Based on the results, we construct a physical model of the
transmission curve and apply it to the data showing that the best-fit model is
consistent with the data within the required calibration accuracy. Finally, in
Section~\ref{sec:discussion}, we show that we can obtain a better fit to the observed
Crab spectrum with the SXS by applying our model with improved parameters.
We summarize the results in Section~\ref{sec:summary}. More details can be found in Midooka (2020)\cite{Midooka20}.

\section{GATE VALVE}
\label{sec:gv} 
\begin{figure}[H]
\begin{center}
	\includegraphics[keepaspectratio, width=0.5\columnwidth]{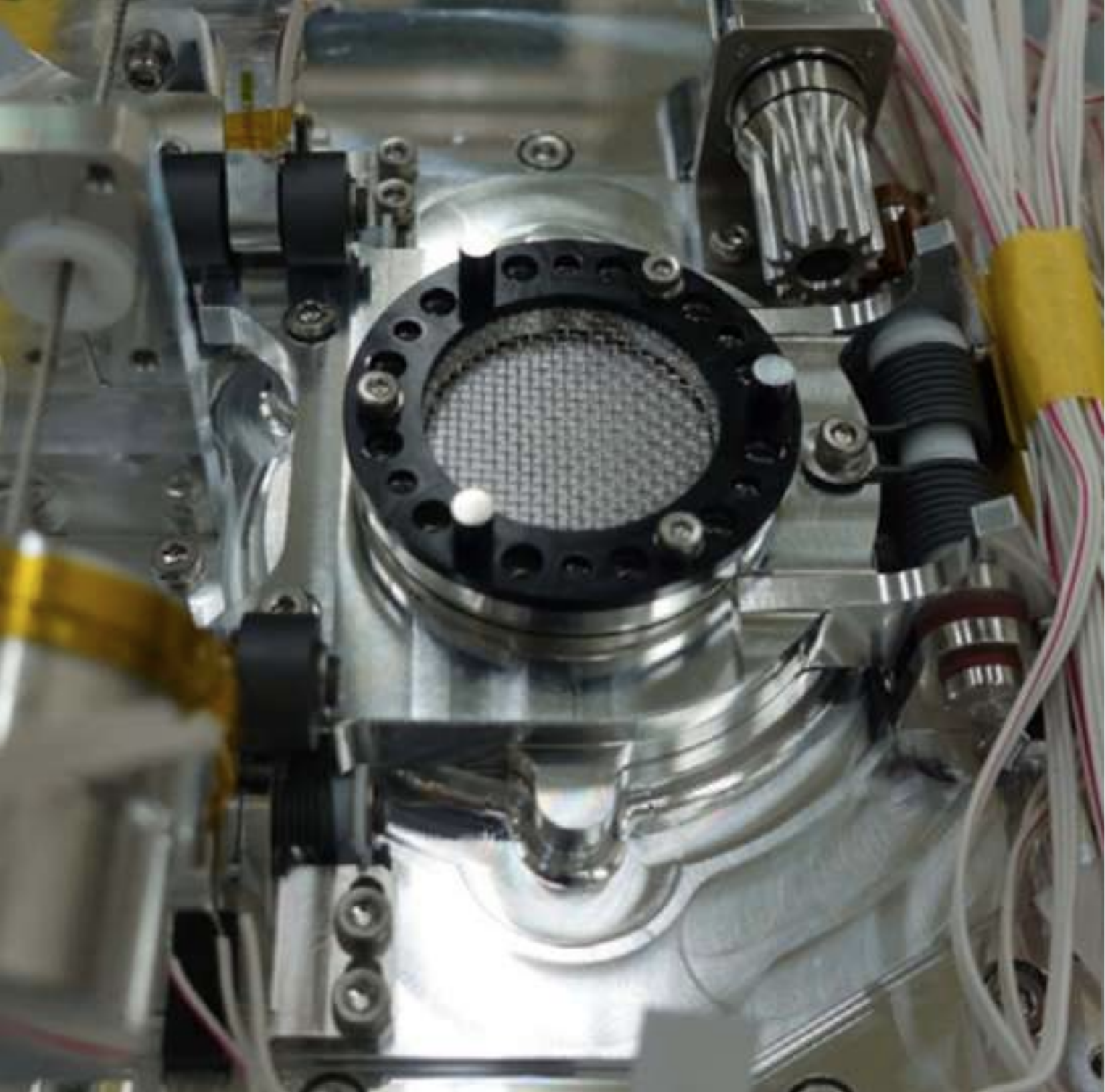}
	    \caption[\textit{Resolve} FM GV at the top of the Dewar]{\textit{Resolve} FM GV in a closed configuration at the top of the Dewar. The photo was provided by Sumitomo Heavy Industries, Ltd.}
	    \label{fig:GV_top_dewar}
\end{center}
\end{figure}

\begin{figure}[H]
\begin{center}
	\includegraphics[width=0.7\columnwidth]{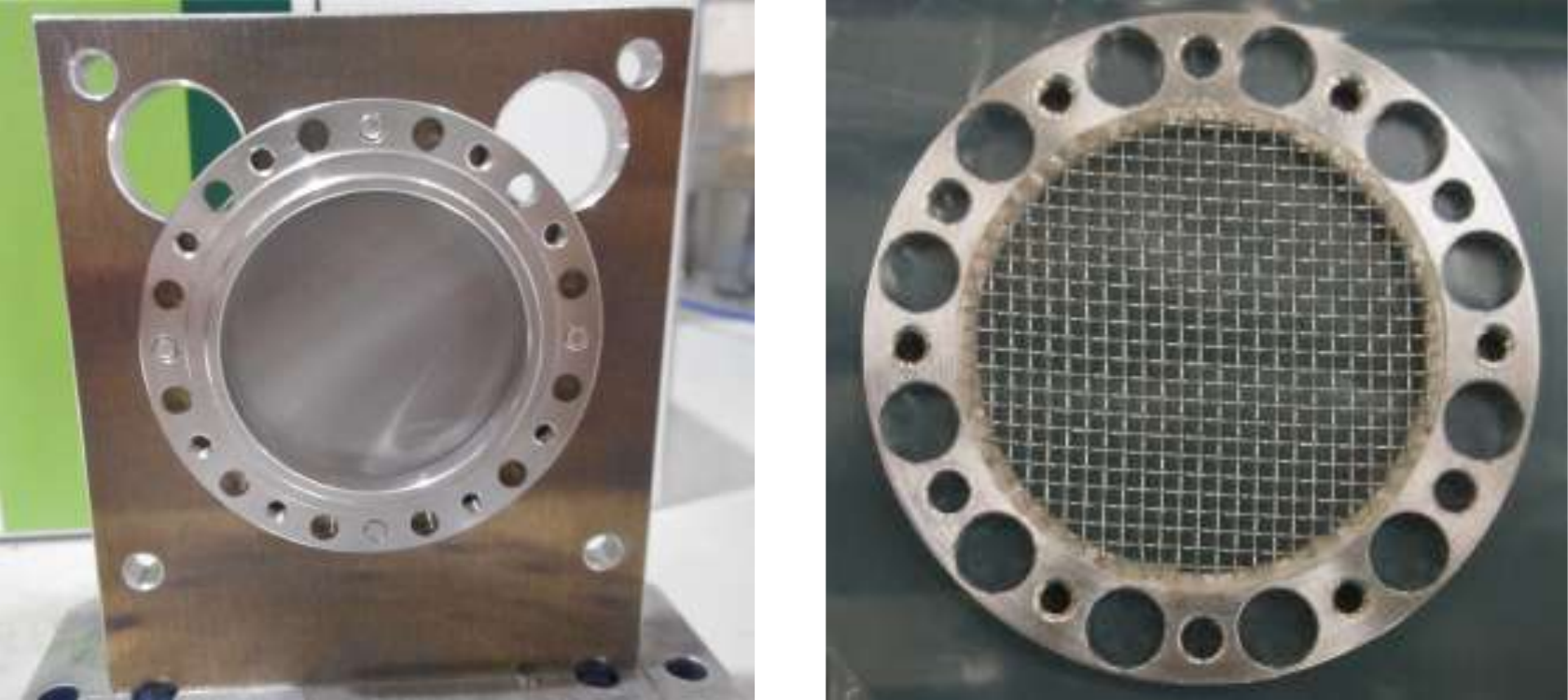}
	    \caption[Picture of the Be window and the stainless steel mesh]{Picture of the Be window (calibration model placed on the holder) and the stainless steel mesh (flight model attached to the flange).}
	    \label{fig:Be_mesh}
\end{center}
\end{figure}

The GV has two parts in the optical path when it is closed (Figure
\ref{fig:GV_top_dewar}): (1) Beryllium window of about 270~\um thickness and (2)
stainless steel mesh for protecting the Be window from accidental damage by any
infalling materials during the ground testing. A picture of these components is shown
in Figure~\ref{fig:Be_mesh}. They have an inner diameter of 29~mm and are placed about
231~mm above the detector.

\paragraph{Be window}
The Beryllium window was fabricated by Materion Corp. from a high purity (IF-1) rolled
foil with a reported thickness of 270$_{-5}^{+8}$~\um. Three models were provided from
the same lot, which we use as the flight model (FM), the flight spare (FS), and the
calibration model (CM). Although the geometrical thickness was measured by the provider,
the X-ray measurements are mandatory because (i) the Be window in a thin film form may
have a lower density than the values in the literature, (ii) the thickness has some
gradient over the diameter in the manufacturing process, and (iii) many spectroscopic
features (absorption edges by atoms of impurities and the features caused by Bragg
diffraction) may affect the high-resolution science by the microcalorimeter. They are
known to vary in different lots from the SXS results \cite{Hoshino17,Yoshida17}.

\paragraph{Stainless steel mesh}
The stainless steel mesh was installed by Sumitomo Heavy Industries, Ltd. The mesh is a
plain weave using the SUS~304 stainless steel wires with a circular cross section. In
the drawing, the wire thickness is 0.20~mm, the pitch is 1.27~mm, and the wire gap is
1.07~mm. The design values are the same for SXS and \textit{Resolve}. From these
values, the opening fraction is estimated to be $(1.07)^2/(1.27)^2 \sim 71\%$. The
remaining 29\% is opaque to the X-rays except for the highest end of the X-ray
bandpass. The mesh for SXS had an optically-thick cross structure at the center for
alignment purposes, which was removed for \textit{Resolve}. The transmission by the
presence of this structure is included in the ray-tracing simulation, which is treated
separately from the average transmission of the mesh part to be presented in this
manuscript. The mesh part was not measured in the SXS.

\section{EXPERIMENTS AND RESULTS} 
\label{sec:results}

\subsection{Energy Dependence of the X-ray Transmission of the Stainless Steel Mesh} 
\begin{figure}[H]
\begin{center}
	\includegraphics[width=1.0\columnwidth]{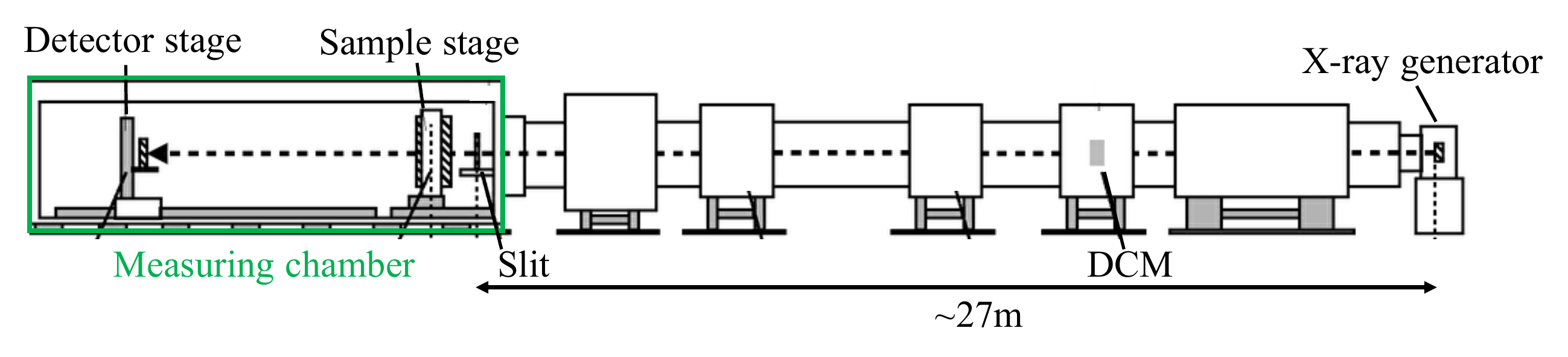}
	    \caption[Schematic of the ISAS beamline]{Schematic of the ISAS beamline \cite{Hayashi15}. The distance between X-ray generator and a slit is approximately 27~m. The slit, sample stage, and detector stage are installed in the measuring chamber.}
	    \label{fig:ISAS_BL}
\end{center}
\end{figure}

We first describe the X-ray transmission measurements of the stainless steel mesh
performed using the 30~m X-ray beamline at JAXA ISAS \cite{Hayashi15} from February 4 to
19, 2019. At that time, the FM stainless steel mesh was not manufactured, thus we used a
spare mesh left from the SXS. Figure~\ref{fig:ISAS_BL} shows the schematic view of the
beamline. X-rays from an X-ray generator illuminate one of the five metal targets: Ti,
Cu, Pt, Mo, and Ag. Their fluorescent and scattered X-rays are further monochromatized
using a Ge (220) double crystal monochromator (DCM). The stainless steel mesh was
installed on the sample stage. The parallel beam was formed by a slit of 2.0 $\times$
2.0~mm, which is comparable in size with the mesh pitch.

\begin{figure}[H]
\begin{center}
	\includegraphics[width=0.7\columnwidth]{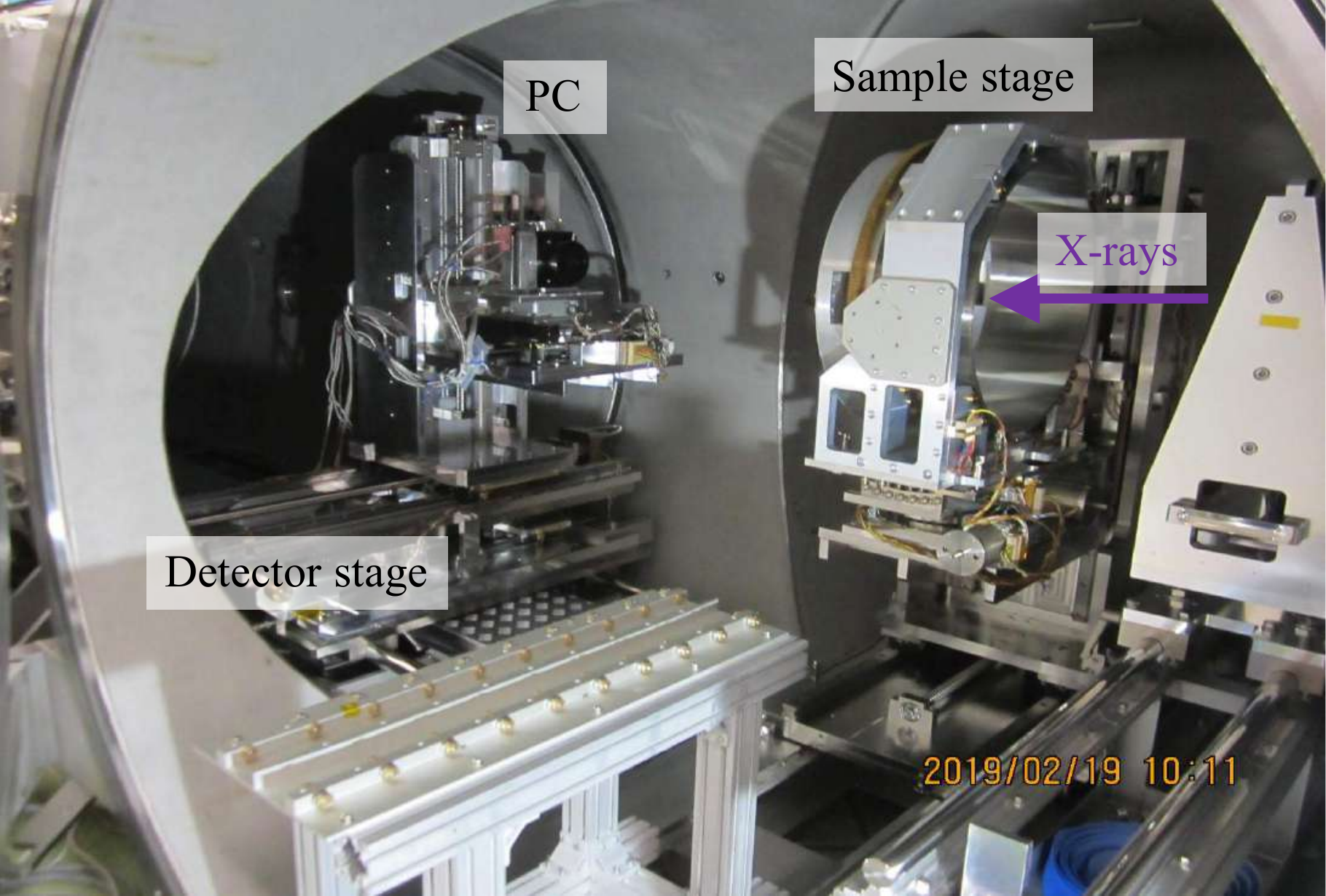}
	    \caption[Interior of the measuring chamber in the beamline]{Interior of the measuring chamber in the beamline.}
	    \label{fig:ISAS_chamber}
\end{center}
\end{figure}

X-ray spectrum with and without the mesh on the sample stage was measured with the
proportional counter (PC) filled with a P10 gas (Ar: 90\%, CH$_4$: 10\%) on the detector
stage. We call them transmitted and direct X-rays, respectively. For the transmitted
X-rays, we performed raster scans over the entire mesh to average for the mesh
pattern. A raster scan is composed of several zigzag lines over entire surface of the mesh by moving the
sample stage continuously at a slow speed of 0.1~mm~s$^{-1}$. The entire surface is
scanned many times to make the systematic error small enough. We performed direct X-ray
measurements before and after the transmitted X-ray measurements without moving the
stages. This set was repeated several times to accumulate sufficient photon statistics. We
then calculated the transmission of the mesh by taking the ratio between the transmitted
and direct X-ray photons in an energy band of interest for each target line. The result
is shown in Table~\ref{tab:meshtable} and the red points with errors in
Figure~\ref{fig:mesh_curve}.  Measurement errors include systematic and statistical
errors, though the latter is negligible. The former was estimated as variations among
individual scans and the latter as Poissonian of the photon counts.

\begingroup
\renewcommand{\arraystretch}{1.2}
\begin{table}[H]
\begin{center}
\caption{Transmission measurement results of the stainless steel mesh}
\scalebox{0.87}{
\begin{threeparttable}
\begin{tabular}{ccccccc}\hline
Target & \begin{tabular}{c}Energy\\\relax[keV] \end{tabular}  & Sample\tnote{a} & \begin{tabular}{c}Total exp.\\\relax [sec] \end{tabular} & \begin{tabular}{c}Total photons\\\relax [counts] \end{tabular}  & \begin{tabular}{c}Transmission\\ \relax[\%] \end{tabular} & \begin{tabular}{c}Error\tnote{b}\\\relax [\%] \end{tabular}\\\hline
\multirow{2}{*}{Ti K{\rm $\alpha$}} & \multirow{2}{*}{4.51} & ON & 1866.3 & 2562596 & \multirow{2}{*}{72.7} & \multirow{2}{*}{$\pm$0.6}\\ 
 & & OFF & 3627.8 & 6850815 & & \\  \hline
\multirow{2}{*}{Cu K{\rm $\alpha$}} & \multirow{2}{*}{8.05} & ON & 1856.7 & 200030 & \multirow{2}{*}{71.7} &  \multirow{2}{*}{$\pm$0.9}\\
 & & OFF & 3601.7 & 540877 & & \\ \hline
\multirow{2}{*}{Pt L{\rm $\beta$}} & \multirow{2}{*}{11.07} & ON & 1858.0 & 422722 & \multirow{2}{*}{71.5} &  \multirow{2}{*}{$\pm$1.5}\\
 & & OFF & 3605.6 & 1149485 & & \\ \hline
\multirow{2}{*}{Mo K{\rm $\alpha$}} & \multirow{2}{*}{17.48} & ON & 3712.8 & 211733 & \multirow{2}{*}{72.3} & \multirow{2}{*}{$\pm$2.3}\\
 & & OFF & 3601.6 & 285442 & & \\ \hline
 \multirow{2}{*}{Mo K{\rm $\beta$}} & \multirow{2}{*}{19.61} & ON & 11136.2 & 122811 & \multirow{2}{*}{73.6} &  \multirow{2}{*}{$\pm$2.8}\\
 & & OFF & 7200.1 & 109824 & & \\ \hline
\multirow{2}{*}{Ag K{\rm $\alpha$}} & \multirow{2}{*}{22.17} & ON & 1861.2 & 270694 & \multirow{2}{*}{76.4} & \multirow{2}{*}{$\pm$1.3}\\
 & & OFF & 3613.6 & 686769 & & \\ \hline
\end{tabular}
	\begin{tablenotes}
		\item[a] \small{Sample ON/OFF means the measurements of the transmitted/direct X-rays.}
		\item[b] \small{The error is for a 1$\sigma$ level.}
	\end{tablenotes}
\end{threeparttable}
}
\label{tab:meshtable}
\end{center}
\end{table}
\endgroup

\subsection{Energy Dependence of the X-ray Transmission of the Be Window} 
\label{sec:E_Be}

X-ray transmission measurements of the FM Be window were performed at KEK Photon Factory
(PF) from June 13 to 23, 2019. We used the beamlines BL-7C\cite{7c} and
BL-11B\cite{11b}. The operating energy ranges are 4.0--12.0~keV and 2.1--5.0~keV and we
used a beam size of 1.0$\times$1.0~mm and 2.5$\times$2.5~mm as the largest possible
square respectively for BL-7C and BL-11B. The intense synchrotron X-rays are
monochromatized using a Si (111) DCM. We measured three positions of the Be window; they
are at the center of the window and at two positions 6.5~mm off-center (top and bottom).

\begin{figure}[H]
 \begin{center}
  \includegraphics[width=0.9\columnwidth]{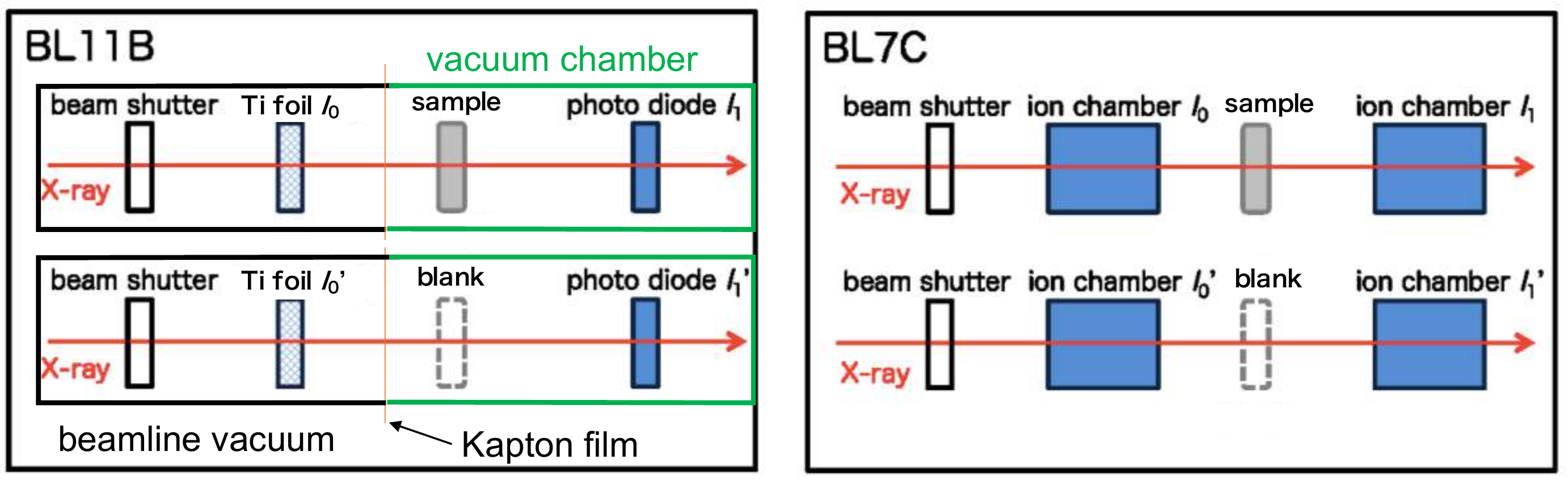}
  \caption[Setup for measurements at the BL-11B and BL-7C]{Schematic drawings of the
  setup for measurements at the BL-11B and BL-7C\cite{Yoshida17}.}
  \label{fig:PF_sche}
 \end{center}
\end{figure}

\begin{figure}[H]
 \begin{center}
  \includegraphics[width=1.0\columnwidth]{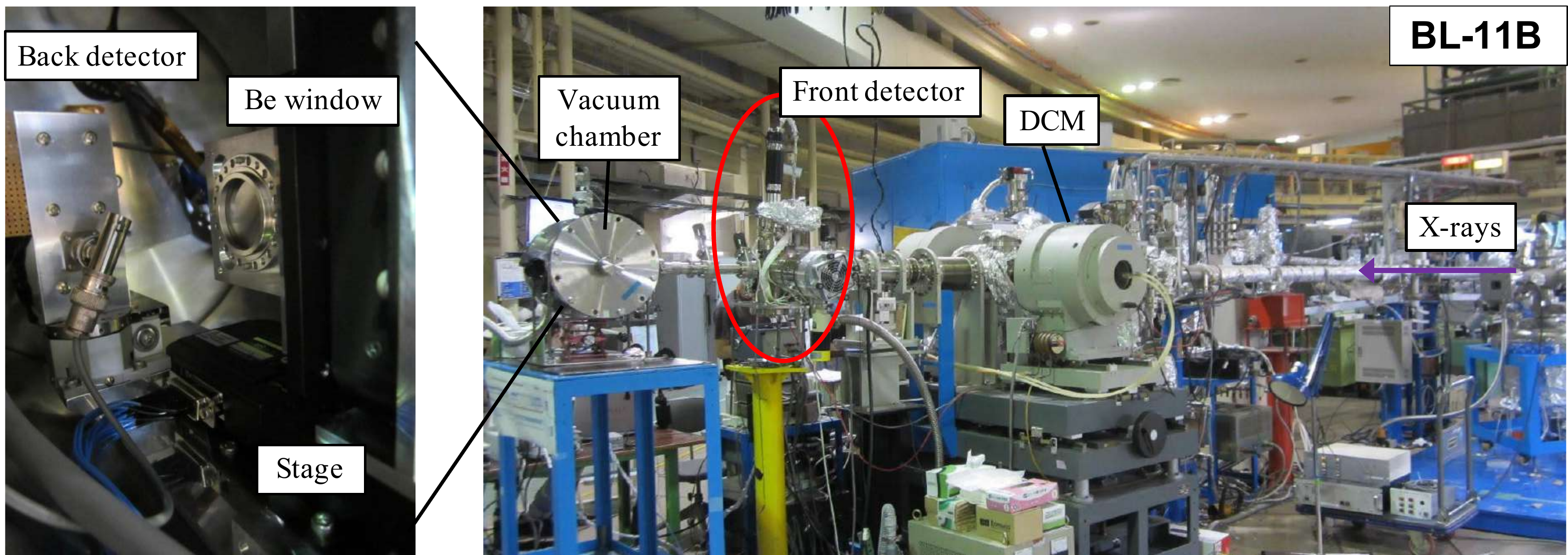}
  \caption[The photo of the BL-11B]{right: The photo of the BL-11B. left:  Interior of
  the vacuum chamber. Be window and a photo diode as the back-side detector are
  installed.}
  \label{fig:BL11B}
 \end{center}
\end{figure}

Figure~\ref{fig:PF_sche} shows the schematic drawings of the setup for measurements at
the BL-11B and BL-7C\cite{Yoshida17}. In the BL-11B experiments, we attached our vacuum
chamber to the beamline through a Kapton polyimide film and installed the Be window and
a photo diode inside the chamber to prevent absorption by the air as shown in
Figure~\ref{fig:BL11B}. The incident X-rays were measured using a Ti foil as a
photocathode beam monitor provided by the facility. In the BL-7C experiments, we
measured incident and transmitted X-rays by two ion chambers filled with N$_2$ gas
provided by the facility in front and back of the Be window placed in the air. 

The contamination by higher-order X-rays are removed by a focusing double mirror in the
BL-7C measurements. The fraction of the remaining third-order X-ray contamination at
4~keV is negligible of about $3 \times 10^{-5}$ \cite{7c}. The contamination has not
been reported in the BL-11B in past experiments at least above 2.3~keV, hence we
consider that they are negligible at the higher energy range in our measurements.

In both experiments, we measured current of the detectors (i.e. the ion chambers, the 
photo-diode, and the Ti foil), as the intensity of the incident and transmitted X-rays.
Here, we assume that $I_0(E)$ and $I_1(E)$ are the
incident and transmitted X-ray intensities with the Be window, and $I'_0(E)$ and
$I'_1(E)$ are those without the Be window at a given energy \textit{E}. The dark current
was measured and subtracted from them. We derived the transmission by
\begin{equation}
T(E)=\frac{I_{1}(E) / I_{0}(E)}{I_{1}^{\prime}(E) / I_{0}^{\prime}(E)}.
\label{eq:trans}
\end{equation}
With this double normalization, we can correct for any instability of the beam as well
as the energy dependence caused by the measurement setup. We repeated this measurement
with a step of 2~eV in eight energy ranges overlapping with each other at 2.35--2.6,
2.55--3.0, 2.95--3.5, 3.45--4.0~keV bands in BL-11B, and 4.0--5.0, 4.9--8.0, 7.9--10.0,
9.9--12.0~keV bands in BL7C. The overlapped range is 50 or 100~eV.

\begin{figure}[H]
 \begin{center}
  \includegraphics[width=1.0\columnwidth]{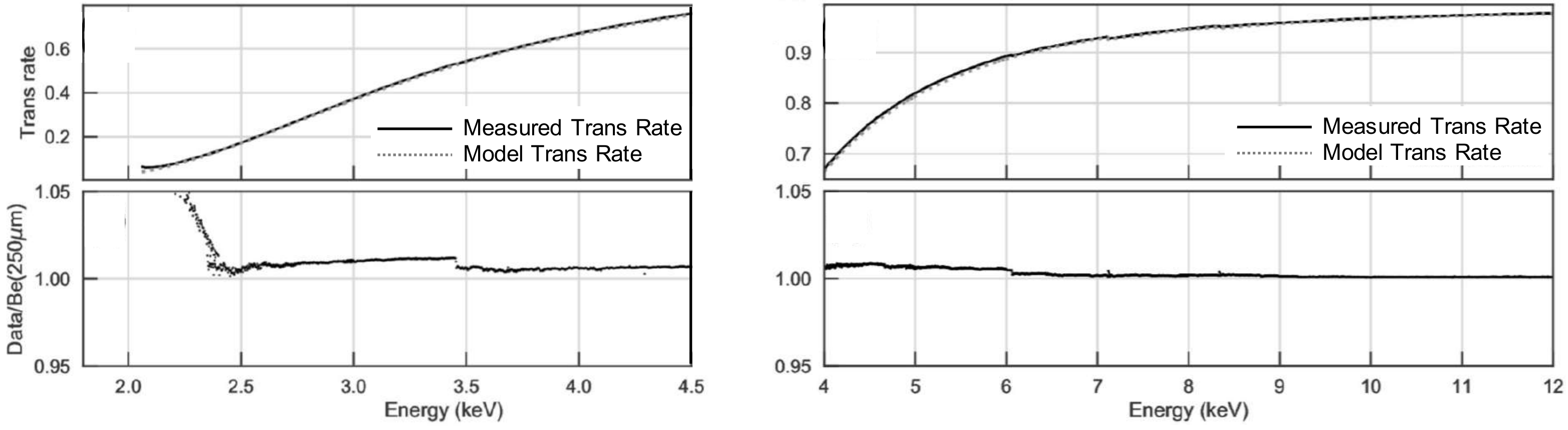}
  \caption[Transmission curve in 1.8--12.0~keV]{Upper panel: A model transmission curve
  assuming a photoelectric absorption by Be of a 250~\um thickness and the transmission
  measured at the top position of the Be window obtained at the BL-11B (left) and BL-7C
  (right). Lower panel: The ratio of the measurement curve to the model curve. The anomalous
  increase of the transmission below 2.4~keV is attributable to higher-order X-ray
  reflections in the DCM.}
  \label{fig:11B_7C}
 \end{center}
\end{figure}

The upper panels of Figure~\ref{fig:11B_7C} show derived transmission curves of the top
position of the Be window and model curves in 2.1--4.5~keV at the BL-11B (left) and in
4.0--12.0~keV at the BL-7C (right). The ratio between two adjacent energy ranges was
corrected by multiplying a constant to the data set in the lower energy of the two. This
was repeated to construct a continuous transmission curve over the entire energy. The
seven constant values provide an estimate of the systematic uncertainty in the
measurements. The mean and the standard deviation of them are 1.00000 and 0.00081,
respectively. We thus evaluate the systematic uncertainty of the measurements to be
0.08\%.

The lower panels of Figure~\ref{fig:11B_7C} show the ratio of the measurement against
the model assuming photoelectric absorption by Be of a 250~\um thickness. Here, we used
the thickness only for displaying purpose to show the residual structures in a limited
range. We found an apparent increase in the transmission below 2.4~keV. After the
delivery of FM Be window, we made an additional measurement using another model (CM) to
investigate this issue at the beamline BL-11 of the Hiroshima Synchrotron Radiation
Center (HiSOR)\cite{hisor} in October 2019. In HiSOR, we examined the contribution of
higher-order X-rays by detuning the two crystal angles in the DCM. The result suggested
some influence of higher-order X-rays in the measurements at KEK PF. We conclude that
the anomalous increase of the transmission below 2.4~keV (Figure.~\ref{fig:11B_7C}) is
likely caused by the effect of higher-order X-ray reflections \cite{Midooka20} and
decided to ignore the data below 2.6~keV in the following analysis.

\subsection{Spatial Dependence of the X-ray Transmission of the Be Window} \label{sec:spatial}
We evaluated the spatial dependence of the transmission using the FM Be window. We used
the beamline BL-11B using almost the same setup in Section~\ref{sec:E_Be}. The beam size
is adjusted to 0.8~mm (horizontal) $\times$ 1.2~mm (vertical).  We measured the X-ray
transmission at some fixed energy over the entire Be window by a scan. We chose four
energies: 2700, 3000, 3456, and 4000~eV. The scan was continuous in the horizontal
direction at a speed of 500~\um~s$^{-1}$ and step-wise in the vertical direction with a
step size of 1~mm. The time series in the horizontal scan was converted to the positions
by the speed and rebinned into a grid map. At the beginning and at the end of each
horizontal scan, we obtained reference transmission without the Be window for 10~s.

\begin{figure}[H]
 \begin{center}
  \includegraphics[width=0.78\columnwidth]{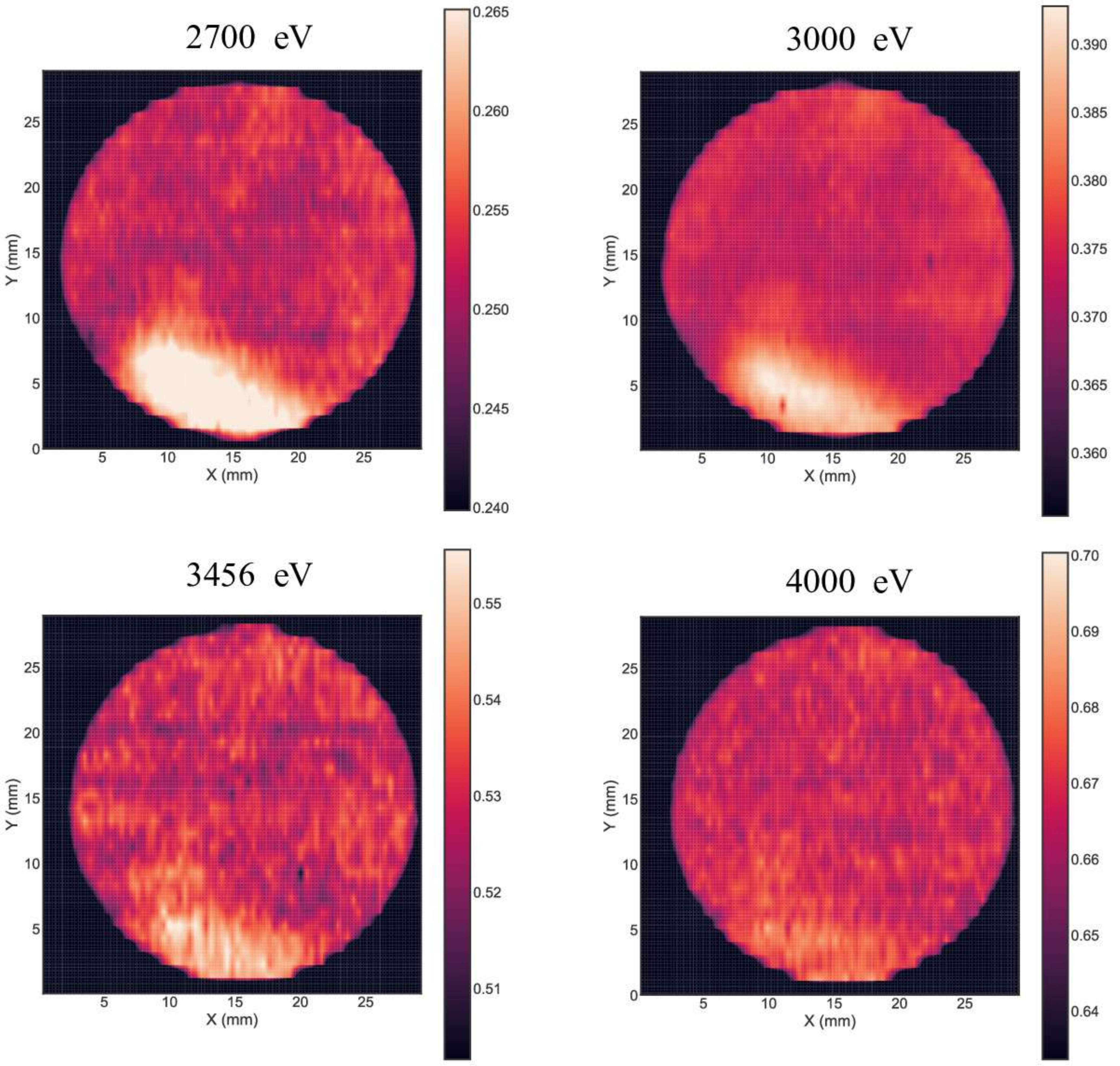}
  \caption[Spatial non-uniformity maps of the Be window transmission]{Spatial
  non-uniformity maps of the transmission of the FM Be window at four selected
  energies. The figures are aligned to the spacecraft coordinate: i.e, $+$X, $+$Y
  directions in the figure are aligned to SAT$+$X  and SAT$+$Y. These maps were faced in
  the direction viewed from the incident X-rays both in the experiment and also to the
  in-orbit setup. The fuzzy boundary is due to an artifact as a result of circular masking.}
  \label{fig:BL11B_raster}
 \end{center}
\end{figure}

Figure~\ref{fig:BL11B_raster} shows the 2D maps of the transmission of the FM Be window.
In all energies, transmission at the lower left of the Be window is higher than the
average value. We speculate that this is caused in the manufacturing process. In the
4000~eV map, the mean and 3\,$\sigma$ scatter of the transmission is
$66.8^{+1.0}_{-1.1}$ \%, which translates to a thickness of 251.5$\pm$10.5~\um. The
non-uniformity is $\pm$4.2\% at 3\,$\sigma$.

\section{DISCUSSION} \label{sec:discussion}

\subsection{Modeling Transmission of the Stainless Steel Mesh} \label{sec:model_mesh}
\begin{figure}[H]
 \begin{center}
  \includegraphics[width=0.8\columnwidth]{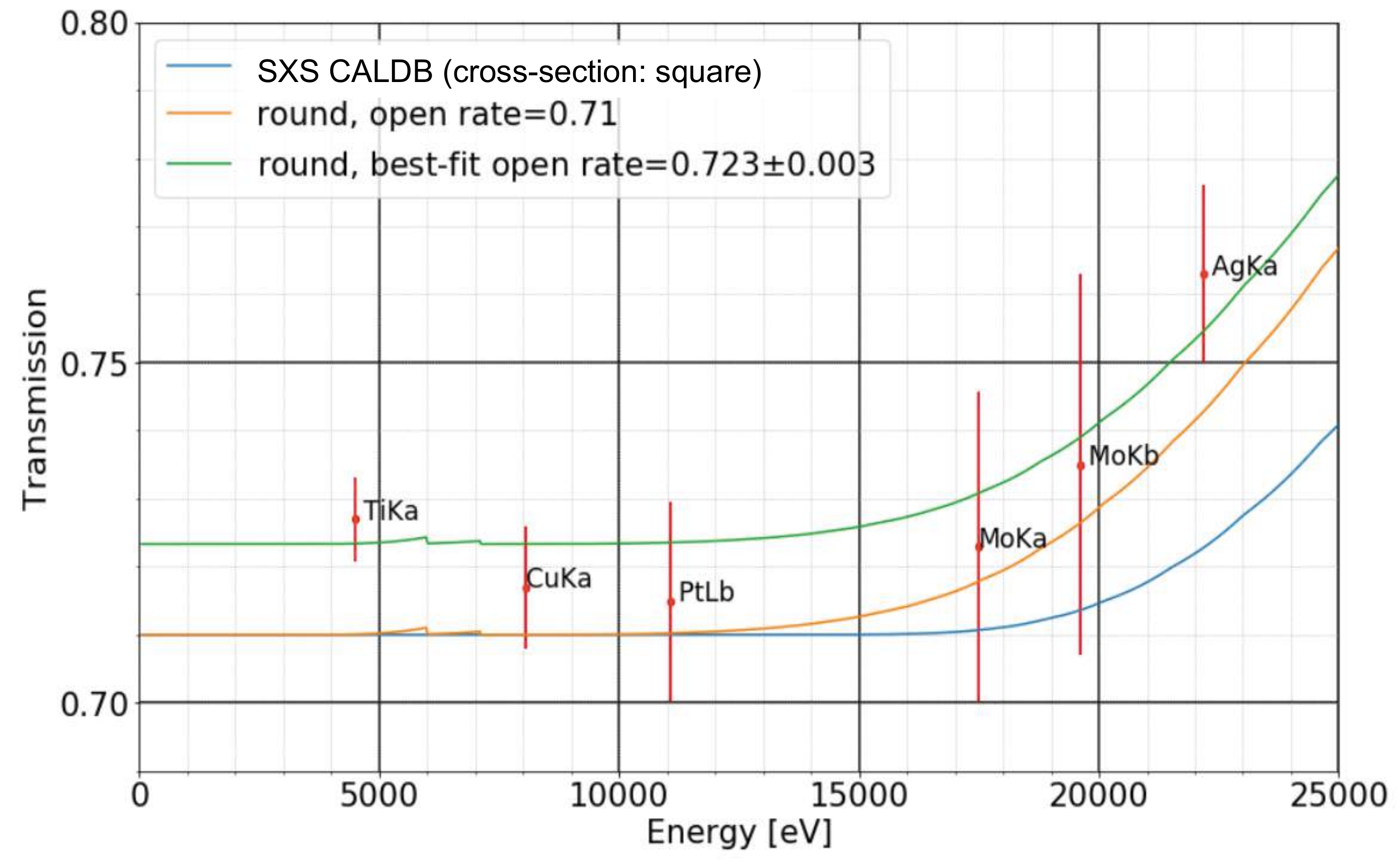}
  \caption[Comparison of three transmission models for the stainless steel
  mesh]{Comparison of three transmission models for the stainless steel mesh. Red points
  describe the measurement results with errors. Blue, orange, green curves show the SXS
  CALDB model, a round cross-section model with a fixed 71\% opening fraction, and the
  \textit{Resolve} model.}
  \label{fig:mesh_curve}
 \end{center}
\end{figure}

In Figure~\ref{fig:mesh_curve}, the measured transmission (Table~\ref{tab:meshtable}) is
shown with the red points with errors, whereas the model curve adopted in the SXS CALDB
is with the blue curve. The mesh is made of SUS~304, which we assumed to have 70\% Fe,
20\% Cr, and 10\% Ni. The transmission is modeled by a product of the photoelectric
absorption and scattering attenuation of these elements.

The model curve adopted in the SXS was found inconsistent with the measurements, in
particular, for the low energy end in Ti K\rmalpha\ (4.51~keV) and the high energy end
in Ag K\rmalpha\ (22.17~keV). One reason is that the cross section of the mesh wire was
assumed a square with 0.2~mm $\times$ 0.2~mm in SXS, while it is round in reality, hence
the actual average depth is smaller than 0.2~mm.  The transmission is non-linear to the
depth, so we performed numerical integration assuming that the wire depth is
$2\sqrt{(t^2/4 - x^2 )}$ at $x (<t/2)$ away from the center of the wire with a diameter
of $t=0.2$~mm. Using the mass attenuation coefficient of a constituent metal
$\mu_\mathrm{tot,metal} (E)$, we derived the total attenuation coefficient $\alpha (E)$
as
\begin{equation}
\alpha(E)=\sum_{\mathrm{metal}} A_{\mathrm{metal}} * \rho_{\mathrm{metal}} * \mu_{\mathrm{tot,metal}}(E),
\end{equation}
where $A_\mathrm{metal}$ is the mass fraction and $\rho_\mathrm{metal}$ is the mass
density of the constituents (Fe, Cr, Ni). The transmission of the stainless steel mesh
$T_\mathrm{mesh}$ ($E$) is then calculated as:
\begin{equation}
T_{\mathrm{mesh}}(E)=f+(1-f) \int_{0}^{\frac{t}{2}} \exp \left(-2 \sqrt{t^{2} / 4-x^{2}} * \alpha(E)\right) d x,
\end{equation}
where \textit{f} is the opening fraction of the mesh. We note that the contribution of
the overlapped parts of the mesh wires can be ignored compared to the measurement error
in Table~\ref{tab:meshtable}.

First, we fixed \textit{f} to 0.71 estimated based on the design in
Section~\ref{sec:gv}, which is shown with the orange curve in
Figure~\ref{fig:mesh_curve}. This is still inconsistent with the
measurements. Considering some uncertainty in the estimate of \textit{f}, we treated it
as a free parameter. The best-fit value was 0.723$\pm$0.003. The resultant model is
shown with the green curve in Figure~\ref{fig:mesh_curve}, which is now consistent with
the measurements. We adopt this model for \textit{Resolve}.

\subsection{Modeling Transmission of the Be Window} \label{sec:model_Be}
First, we investigated contamination by impurities. In the lower panel of Figure~\ref{fig:11B_7C}, we identified features associated with the photoelectric absorption edges of Cr-K, Mn-K, Fe-K, Ni-K, and Cu-K, but not others. We derived their areal densities by local fitting and summarize the result in Table~\ref{tab:minor_thick}.
The result is compared with the report by Materion Corp.

\begin{table}[H]	
\caption[Areal density of impurities of the FM Be window]{Best-fit areal density of the impurities based on the transmission measurements of the FM Be window.}	
\label{tab:minor_thick}	
\begin{center}	
\begin{threeparttable}	
\begin{tabular}[t]{cccccc}\hline	
Impurity  & Edge energy & Fitting energy & Best-fit areal\tnote{a} & Reference areal\tnote{b}\\	
element & [eV] & range [eV] &  density [\ug/cm$^2$] &  density [\ug/cm$^2$]\\\hline	
Cr & 5989 & 5500--6050 & 0.8 $\pm$ 0.1 & 1\\	
Mn & 6539 & 6510--6900 & $<$ 0.2 & 1\\	
Fe & 7112 & 7000--7500 & 11.4 $\pm$ 0.2 & 9\\	
Ni & 8333 & 8200--8450 & 7.7 $\pm$ 0.2 & 5\\	
Cu & 8979 & 8800--9180 & 0.7 $\pm$ 0.1 & 1\\\hline	
\end{tabular}	
\begin{tablenotes}	
\item[a] \small{The uncertainties are at the 1$\sigma$ confidence level.}	
\item[b] \small{The values are derived from the fraction of materials measurements by Materion Corp.}	
\end{tablenotes}	
\end{threeparttable}	
\end{center}	
\end{table}

We also detected other edge-like features at 3460, 6057~eV, and so on. They are physically interpreted as the Bragg diffraction in the Be window \cite{Yoshida17,Hoshino17}. Here, the Be window is a poly-crystal, in which a part of the material satisfies the Bragg condition at any given energy. An ideal case is called the powder diffraction and the diffracted X-rays form a ring called the Debye-Scherrer ring. Hereafter, we call these edge-like features as Bragg diffraction features (BDFs). 
When the incident X-ray energy satisfies the Bragg condition expressed in
Equation~(\ref{eq:bragg}) for the lattice spacing $d_{hkl}$ of the Miller index ($hkl$),
X-rays are coherently scattered into a Debye-Scherrer ring. The ring disappears when the
incident energy becomes lower than the BDF energy $E_{hkl}(\theta = \pi/2)$, as the
scattered angle cannot exceed $2\theta = 180^{\circ}$ . This makes a discontinuity in
the measured transmission curve at $E_{hkl}(\theta = \pi/2)$ of all Miller indices
except for those with systematic absence (Appendix~\ref{sec:appendix}).
\begin{equation}
E_{hkl}(\theta)=\frac{h c}{2 d_{hkl} \sin \theta},
\label{eq:bragg}
\end{equation}
where $\theta$ is the diffraction angle, $h$ is the Planck constant, and $c$ is the speed of light.

\begin{figure}[H]
\begin{center}
\includegraphics[keepaspectratio, scale=0.6]{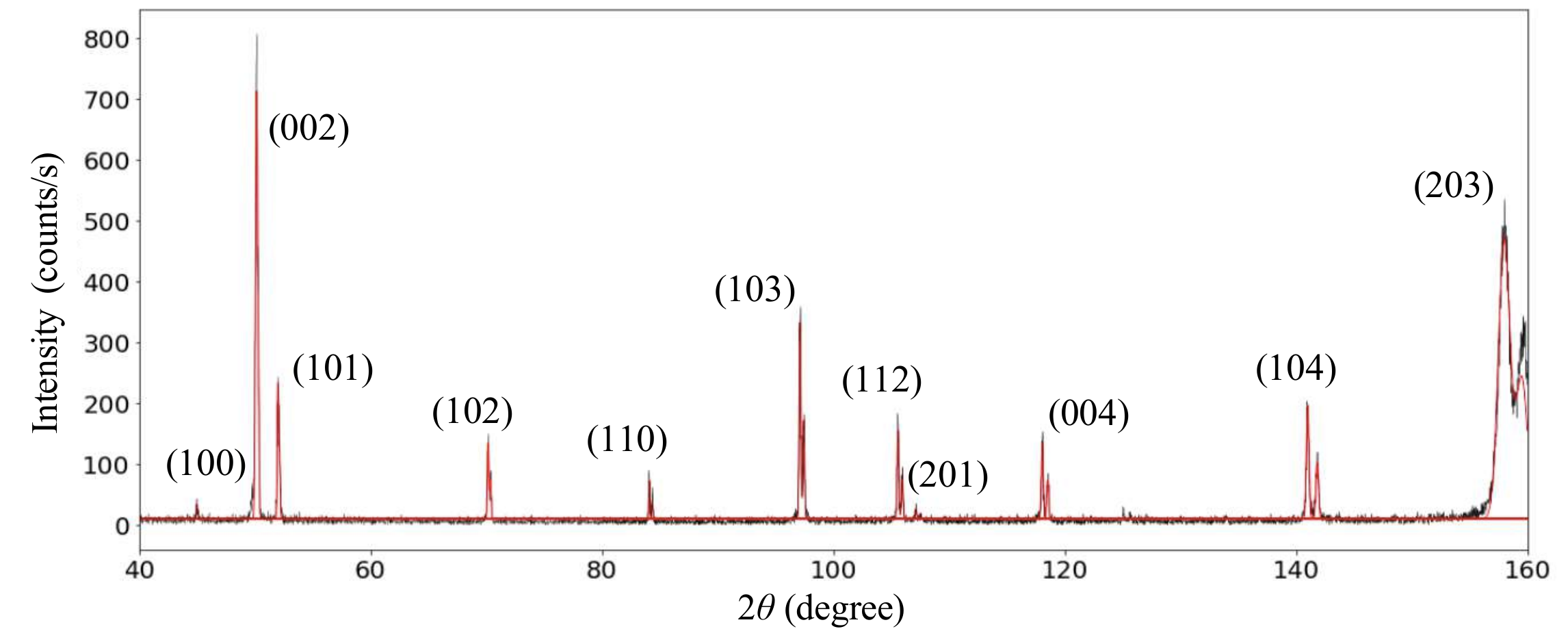}
\caption[Results of the X-ray diffraction measurements and fitting]{Results of the X-ray diffraction measurements and fitting. All peaks are identified with a Miller index at an expected diffraction angle. All peaks have double-peak structures due to mixing of the Cu K\rmalpha$^1$ and K\rmalpha$^2$ lines in the incident X-rays.}
\label{fig:xrd_fit}
\end{center}
\end{figure}

The interpretation was confirmed in an additional measurement using CM. We performed an
X-ray diffraction measurement using a multi-purpose X-ray diffractometer (Rigaku Ultima
IV) in the Tokyo Metropolitan University. Figure~\ref{fig:xrd_fit} shows the intensity
of the scattered Cu K$\alpha$ line X-rays as a function of the diffraction angle. All
peaks are identified with a Miller index and correspond to a Debye-Scherrer ring. The
peak heights in the diffraction is in an agreement with the edge depths in the
transmission for each Miller index. The result endorses our interpretation that the
features in the transmission curve are indeed the BDFs.

\begin{figure}[H]
\begin{center}
    \includegraphics[width=0.9\columnwidth]{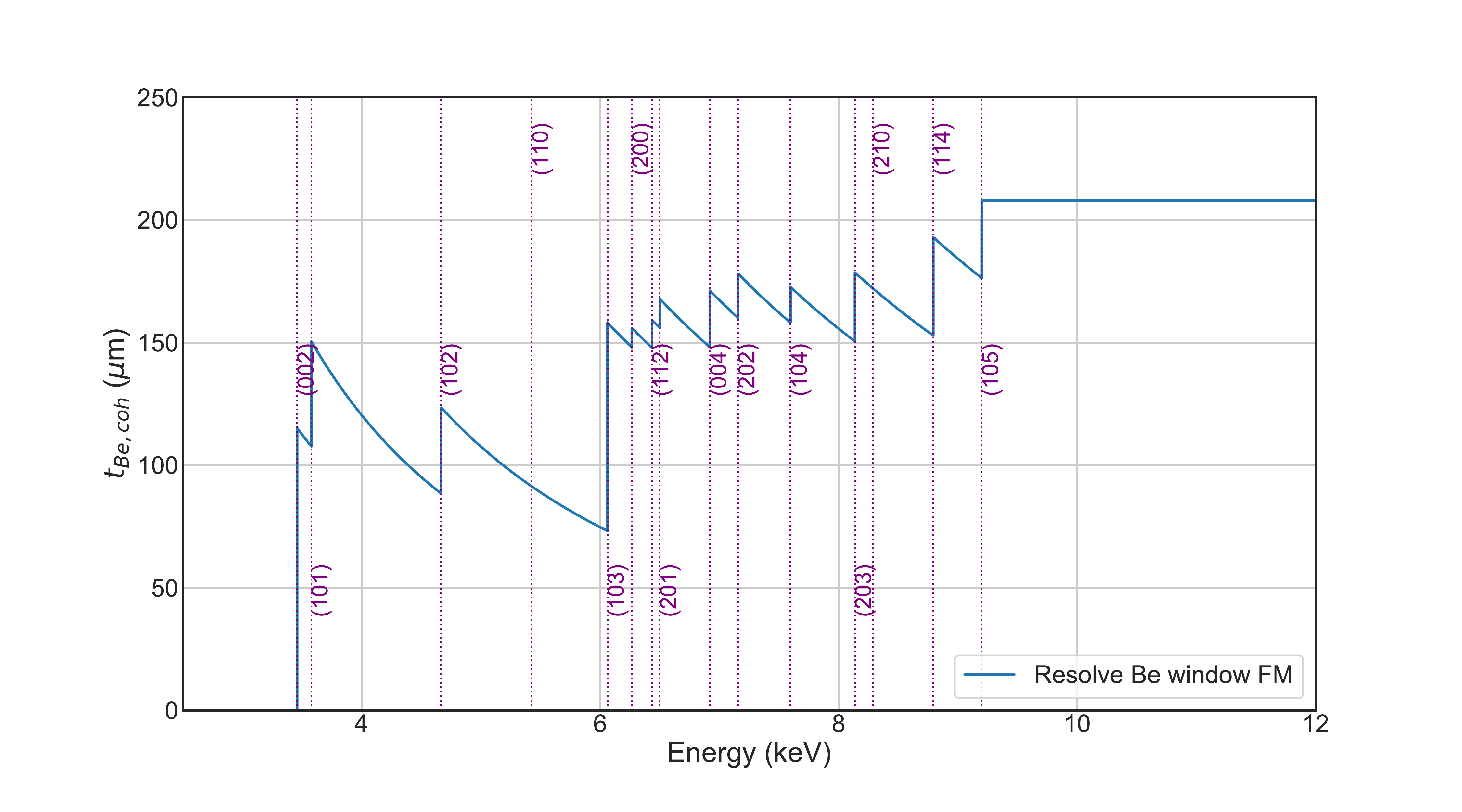}	
    \caption[Best-fit coherent scattering thickness of the Be with step-wise
 changes]{Best-fit thickness of coherent scattering of Be with step-wise changes. Purple
 lines represent the BDF energies apparent in the data, and the numbers in parentheses
 represent the crystal planes $(hkl)$ that cause the BDF. We summarize the actual
 $t_{\mathrm{Be, coh}}(E)$ values in Table~\ref{tab:coh_depth}.}
	\label{fig:thick_coh}
\end{center}
\end{figure}

We took the following approach in modeling the BDFs. We used the thickness of the photoelectric absorption, coherent scattering, and incoherent scattering as phenomenological fitting parameters.	
The energy dependence of the mass attenuation coefficients for the three interactions was fixed to the values by the NIST database\cite{xcom}. This approach was motivated by the fact that the scattering cross section effectively changes in lattice from the values in atoms in dynamical diffraction theories \cite{Authier12}. Considering that the BDF is caused by the energy dependence of the anisotropy of coherent scattering, we changed the coherent scattering thickness in a step-wise manner at all BDFs. In between two BDFs, we assumed that the thickness is proportional to the inverse square of the incident energy (Figure~\ref{fig:thick_coh}). This is derived from the discussion\cite{oxford} on the cross section of the powder diffraction as	

\begin{equation}	
\label{e05}	
\frac{d\sigma}{d\Omega} = N_{c} \left(\frac{d_{\tau}^3}{v_0}\right) M_{\tau}|F(\tau)|^{2}\frac{\tan{\theta}}{2\pi} \delta(\gamma),	
\end{equation}	
in which $N_c$ is the number of unit cells in the powder crystal, $v_0$ is the volume of
a unit cell, $\tau$ is the reciprocal vector corresponding to a Miller index, $F(\tau)$
and $M_{\tau}$ are respectively the structure factor and the reflection multiplexity
factor for the reflection $\tau$, and $d_{\tau}$ is the lattice spacing. The Bragg condition is met at $hc/E = 2d_{\tau}\sin{\theta}$. The angle $\gamma=2\theta$ and $d\Omega = 2\pi \sin{\gamma}d\gamma$.	
The Equation~(\ref{e05}) is derived for the constant-wave case, but it applies to the
constant-energy case as well as wavelength-dispersive case with variable
conversions. The scattered photon counts per second integrated over the entire solid
angle (or entire energy range through the Bragg condition) is then given by	
\begin{equation}	
P_{\tau}(E) = \int dE' \frac{d\sigma}{d\Omega}\frac{d\Omega}{dE'} \Phi(E') = N_{c} \left(\frac{d_{\tau}}{v_0}\right) M_{\tau}|F(\tau)|^{2}\Phi(E)\frac{h^2c^2}{4E^2},	
\end{equation}	
in which $\Phi(E)$ is the incident photon counts per second per unit area per energy
bin. From these, $P_{\tau}(E)/\Phi(E) \propto E^{-2}$, which translates to the thickness
dependence on $E$ when the scattering depth is optically thin. The thickness was left
constant against energy beyond $\sim$9.2~keV, where there are numerous BDFs and we
observed no discontinuous structures in the data.


Table~\ref{tab:coh_depth} shows best-fit coherent scattering thickness of the
\textit{Resolve} FM Be window, which accounts for BDFs.  Considering these values and
the dependency of $E^{-2}$, we made the plot of the blue curve in Figure~\ref{fig:thick_coh}.

\begin{table}[H]
\begin{center}
\caption[Fitting results of the coherent scattering thickness of the \textit{Resolve} FM
 Be window]{Fitting results of the coherent scattering thickness at each BDF for the
 \textit{Resolve} FM Be window.}
\label{tab:coh_depth}
\begin{threeparttable}
\scalebox{0.8}{
\begin{tabular}[t]{ccccc}\hline
\begin{tabular}{c}Miller index\\\relax $hkl$\end{tabular} & \begin{tabular}{c}BDF energy\\\relax [keV]\end{tabular} & \begin{tabular}{c}Fitting range\\\relax [keV]\end{tabular} & \begin{tabular}{c}Coherent scattering \\\relax thickness $t_{\rm Be, coh}$ [\textmu m]\end{tabular} & \begin{tabular}{c}$t_{\rm Be, coh}$ difference \tnote{b}\\\relax above/below BDF energy [\textmu m]\end{tabular}\\\hline
100 & 3.1326  & N/A & 0\tnote{a} & 0\\
002 & 3.4596  & 3.47--3.57 & 115.2$\pm$1.6 & 115.2\\
101 & 3.5785  & 3.59--4.66 & 150.5$\pm$0.8 & 42.8\\
102 & 4.6671  & 4.67--5.42 & 123.5$\pm$0.8 & 35.0\\
110 & 5.4259  & 5.43--6.05 & 88.5$\pm$0.7 & 0.02\\
103 & 6.0616  & 6.07--6.25 & 158.2$\pm$1.1 & 85.0\\
200 & 6.2653  & 6.27--6.43 & 156.0$\pm$0.9 & 7.9\\
112 & 6.4350  & 6.44--6.49 & 159.4$\pm$1.4 & 11.3\\
201 & 6.4999  & 6.51--6.91 & 167.9$\pm$0.6 & 11.9\\
004 & 6.9191  & 6.93--7.1 & 171.2$\pm$0.6 & 23.0\\
202 & 7.1569  & 7.16--7.59 & 178.0$\pm$0.5 & 18.1\\
104 & 7.5953  & 7.6--8.13 & 172.7$\pm$0.4 & 14.6\\
203 & 8.1353  & 8.14--8.78 & 178.6$\pm$0.9 & 28.1\\
114 & 8.7929  & 8.8--9.19 & 192.9$\pm$0.5 & 40.1\\
105 & 9.1988  & 9.2--12.0 & 208.0$\pm$0.2 & 31.7\\\hline
\end{tabular}
}
 \begin{tablenotes}
  \item[a] We set $t_{\rm Be, coh}=0$ below the lowest BDF energy at the (002) plane.
  \item[b] These values are used to plot blue curve in Figure~\ref{fig:thick_coh}.
 \end{tablenotes}
\end{threeparttable}
\end{center}
\end{table}

Our model consists of the following components:
\begin{itemize}
    \item Photoelectric absorption and incoherent scattering by the bulk Be to account for the overall shape of the transmission curve.
    \item Photoelectric absorption by impurities (Cr, Mn, Fe, Ni, and Cu) to account for the absorption edges.
    \item Coherent scattering with step-wise changes in thickness to account for the BDFs of Be poly-crystals.
\end{itemize}

The total transmission is given by 
\begin{equation}
T_{\mathrm{Be}}(E)=T_{\mathrm{Be, photo+inc}}(E) * T_{\mathrm{imp, photo}}(E) * T_{\mathrm{BDF}}(E),
\label{eq:three_models}
\end{equation}
where $T_\mathrm{Be,photo+inc} (E)$ is the transmission due to the photoelectric absorption and incoherent scattering by the bulk Be expressed as:
\begin{equation}
T_{\mathrm{Be, photo+inc}}(E)=\exp \{-t_{\mathrm{Be, photo+inc}} * \rho_{\mathrm{Be}} * (\mu_{\mathrm{Be, photo}}(E) + \mu_{\mathrm{Be, inc}}(E)) \}.
\label{eq:photoinc}
\end{equation}
Here, $t_\mathrm{Be,photo+inc}$ is the thickness of the bulk Be common to the photoelectric absorption and the incoherent scattering. $\rho_\mathrm{Be}$ is the Be density. $\mu_\mathrm{Be,photo} (E)$ and $\mu_\mathrm{Be,inc} (E)$ are, respectively, the mass attenuation coefficient of photoelectric absorption and incoherent scattering as a function of incident X-ray energy $E$.
$T_\mathrm{imp,photo} (E)$ is the transmission due to the photoelectric absorption by the impurities, which is calculated by
\begin{equation}
T_{\mathrm{imp,photo}}(E)=\prod_{\mathrm{imp}} \exp \{-\sigma_{\mathrm{imp}} * \mu_{\mathrm{imp,photo}}(E)\},
\label{eq:imp}
\end{equation}
where $\sigma_{\mathrm{imp}}$ is the areal density of the impurities present in the bulk Be. We modeled the transmission of the BDFs by
\begin{equation}
T_{\mathrm{BDF}}(E)=\exp \{-t_{\mathrm{Be, coh}}(E) * \rho_{\mathrm{Be}} * \mu_{\mathrm{Be, coh}}(E)\}.
\label{eq:bdf}
\end{equation}

\begin{figure}[H]
\begin{center}
	\includegraphics[width=0.8\columnwidth]{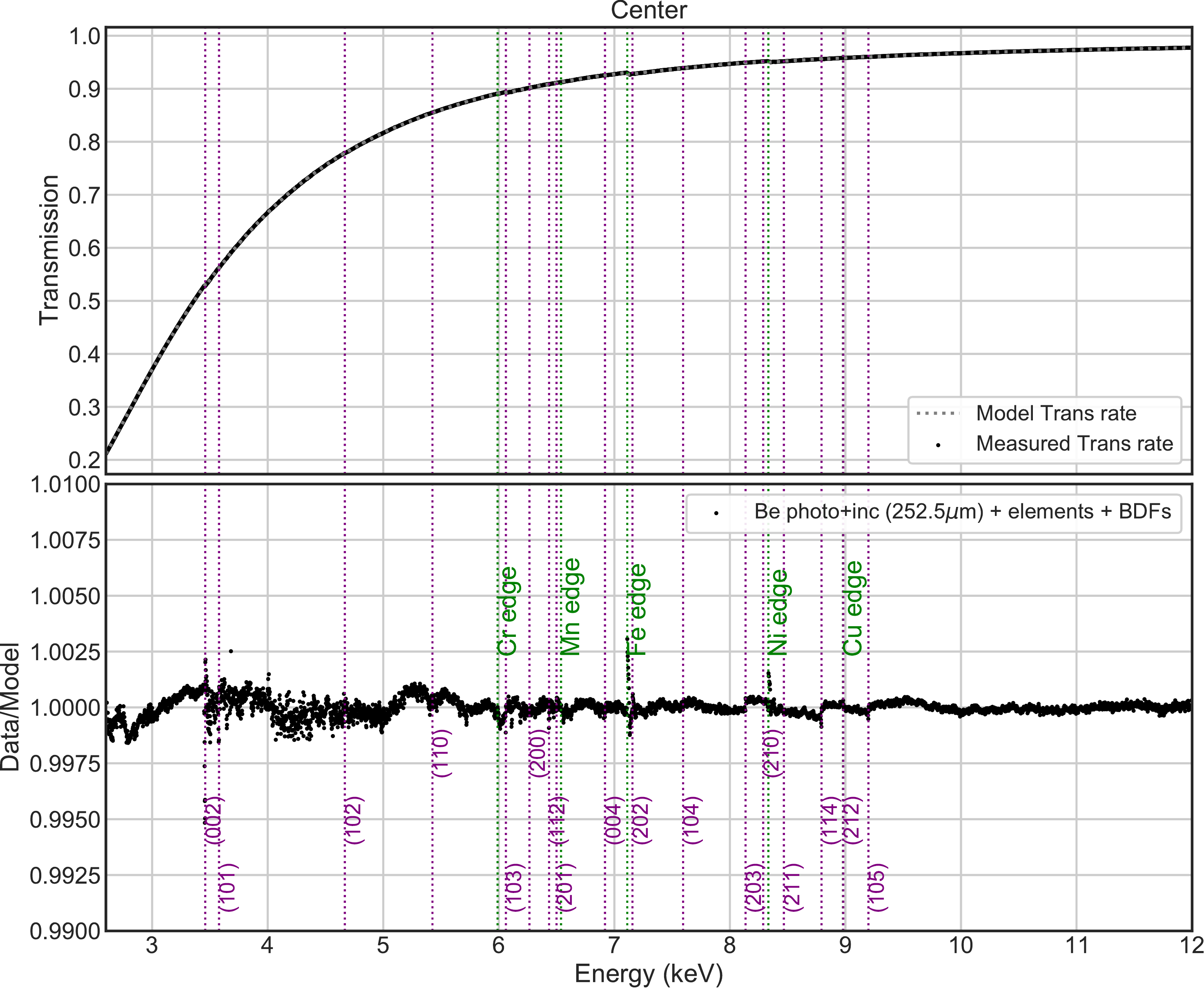}
	    \caption[Data and best-fit transmission model of the FM Be window in 2.6--12.0~keV]{Data and best-fit transmission model of the FM Be window in 2.6--12.0~keV (top). The difference between the data and the model (bottom) is smaller than $\sim$0.2\% in the entire measured energy range. The edges by the contaminants are shown in green, while the BDFs of the labeled Miller index is in purple.}
	    \label{fig:Be_best}
\end{center}
\end{figure}

Figure~\ref{fig:Be_best} shows the best-fit model for the \textit{Resolve} FM Be
window. 
The best-fit parameters are: $T_{\mathrm{Be, photo+inc}}(E)=252.5$ \um for Equation~\ref{eq:photoinc}, values in Table~\ref{tab:minor_thick} for Equation~\ref{eq:imp}, and those in Table~\ref{tab:coh_depth} for Equation~\ref{eq:bdf}.
The difference between the data and the model is smaller than $\sim$0.3\% over a
wide energy range, which satisfies the calibration requirements described in
Section~\ref{sec:intro}.  The mass attenuation coefficients by NIST have some
uncertainties. However, they are accounted to some extent by being multiplied by a free
parameter. In total, we estimate the transmission uncertainty smaller than $\sim$0.3\%
in 2.6--12.0~keV.

As shown in Figure~\ref{fig:BL11B_raster}, the Be thickness has a spatial non-uniformity
of $\pm 4.2\%$. We plan to include this in CALDB as a spatial map. Other spatial
differences, such as the absorption edges by the impurities and the BDFs, were found
small enough not to affect the observation results based on the comparison of the
measurements at three positions (center, top, and bottom) of the Be window.

\begin{figure}[H]
\begin{center}
 \includegraphics[width=0.8\columnwidth]{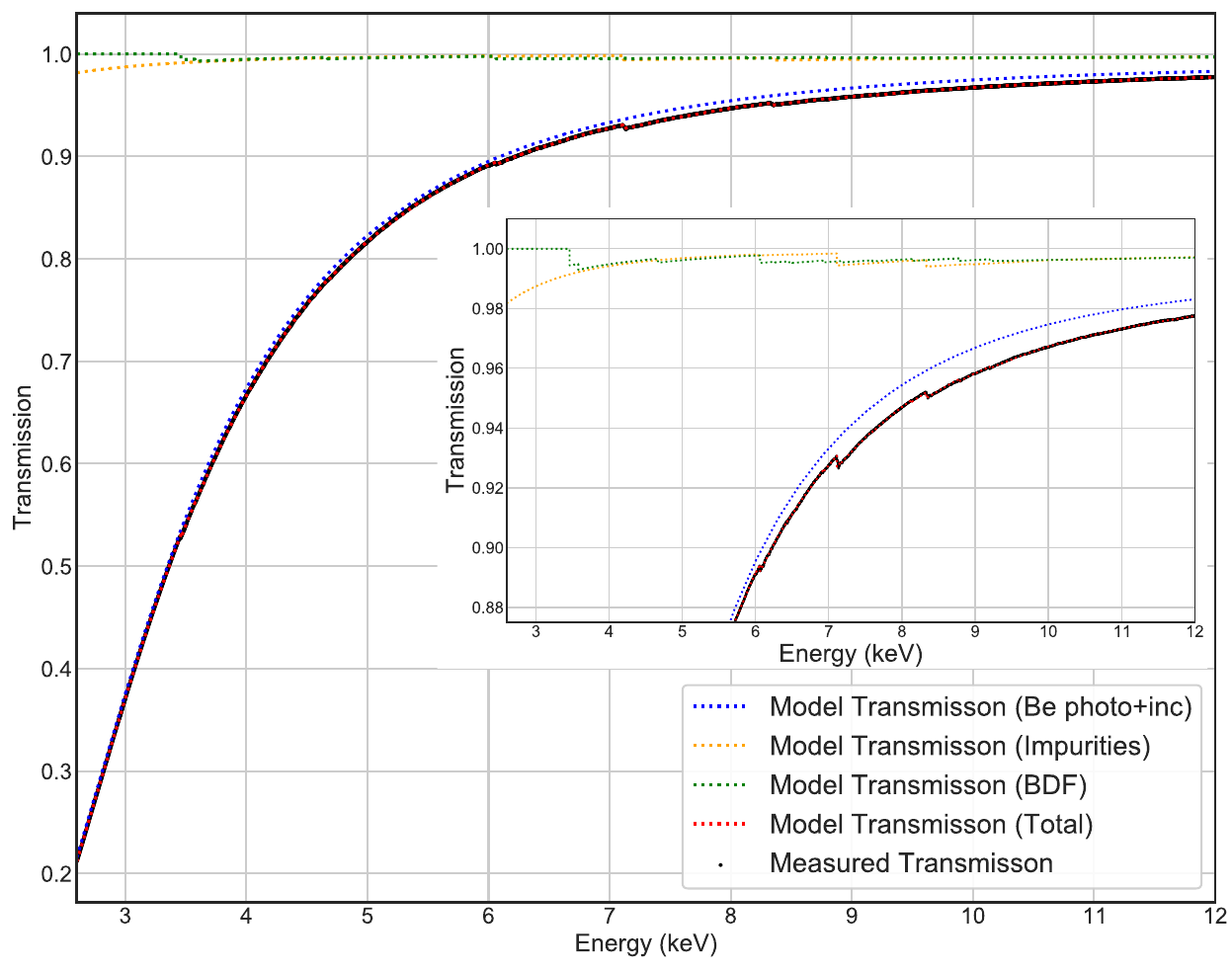}
 \caption[Decomposition of the three transmission model components]{Comparison of the three transmission model components in Equation~(\ref{eq:three_models}). Inset is an enlarged view of the area where the y-axis is greater than 0.875.}
 \label{fig:eachmodel}
\end{center}
\end{figure}

Figure~\ref{fig:eachmodel} shows the three transmission components expressed in
Equation~(\ref{eq:three_models}). Photoelectric absorption and incoherent scattering of
Be (blue) dominates the overall transmission curve. The BDFs (green) and photoelectric
absorption of impurities (orange) account for local edge-like features expanded in the
inset.

The BDF model in this study is quite similar with that developed for SXS by Yoshida et
al. (2017)\cite{Yoshida17}. Both have step-wise discontinuities followed by a $E^{-2}$
dependency on photon energy $E$. We gave some physical explanation for this behavior and
could impose some constraints global to the energy range. Together with improved data
quality, we could better constrain the model parameters now.

\subsection{Modeling Total Transmission of the GV} 
We derived the best-fit model for the stainless steel mesh $T_\mathrm{mesh} (E)$ in
Section~\ref{sec:model_mesh} and the Be window $T_\mathrm{Be} (E)$ in
Section~\ref{sec:model_Be}. We finally multiply them to obtain the total transmission
curve of the GV $T (E)$ as
\begin{equation}
T(E) = T_{\mathrm{Be}}(E)\left\{f+(1-f) T_{\mathrm{mesh}}(E)\right\},
\end{equation}
where \textit{f} is fixed to the best-fit value of 0.723. The model was extrapolated to cover the full range from 10 to 40000~eV and interpolated to fill in the 0.25~eV pitch to deliver as a \textit{Resolve} CALDB product.



The ground calibration result presented here has some limitations, which need to be
verified with in-orbit observations. We will (1) evaluate any systematics caused by
differences in the measurement setup and  actual observation such as beam angles, the
detector solid angle, etc, and  (2) check the validity of the linear extrapolation of
the transmission model out of the measured energy range. We have several free parameters
in our model. Fine-tuning these parameters will be achieved by observing continuum
sources such as pulsar wind nebulae and blazers in reasonable telescope times before and
after the GV open \cite{Midooka20, Miller20}.

\subsection{Revisit of Crab Spectrum Obtained with Hitomi SXS} 
We have so far discussed the transmission modeling in the \textit{Resolve} GV. Our
approach in the GV modeling can be applied to the ground measurement data taken for the
SXS and be compared to the observed data with the SXS in orbit. For the stainless
steel mesh, no measurements were done for the SXS, so we fixed the opening fraction
$f=0.71$ based on the drawings. This is because, as seen in Figure~\ref{fig:mesh_curve},
the mesh is optically thick in the spectral fitting range 2--12~keV and the influence on
the residuals is small. For the Be window, we used the ground measurement results of the
SXS FS model \cite{Yoshida17} obtained in the same setup with Section~\ref{sec:E_Be} and
applied our modeling in Section~\ref{sec:model_Be}. We derived the best-fit Be
photoelectric absorption thickness to be 272.5~\um.

We generated a Crab spectrum based on the IACHEC model of the Crab nebula established
for cross-calibration purposes\cite{Weisskopf10}. The model is a simple power-law of a
photon index 2.10 and a normalization of 8.70 attenuated by an interstellar absorption
of an equivalent hydrogen column density of $0.42\times10^{22}$~cm$^{-2}$. We
convolved the model with the redistribution matrix function (RMF) of the SXS CALDB and
an ancillary response file (ARF) and compared it to the observed spectrum. We modified
the ARFs by allowing some deviations in the Be window 
thickness. This is because that the FM Be window used in orbit is different from the
FS Be window used in the ground measurement and that the difference can be as large as
the level of spatial non-uniformity found in the \textit{Resolve} FM Be window of
$\pm$4.2\% (Section~\ref{sec:spatial}). We made a comparison with the data for the
modified ARFs of a Be thicknesses of 261, 266, 271, 276, and 281~\um and computed the
$\chi^{2}$ values.

\begin{figure}[H]
 \begin{center}
  \includegraphics[width=0.8\columnwidth]{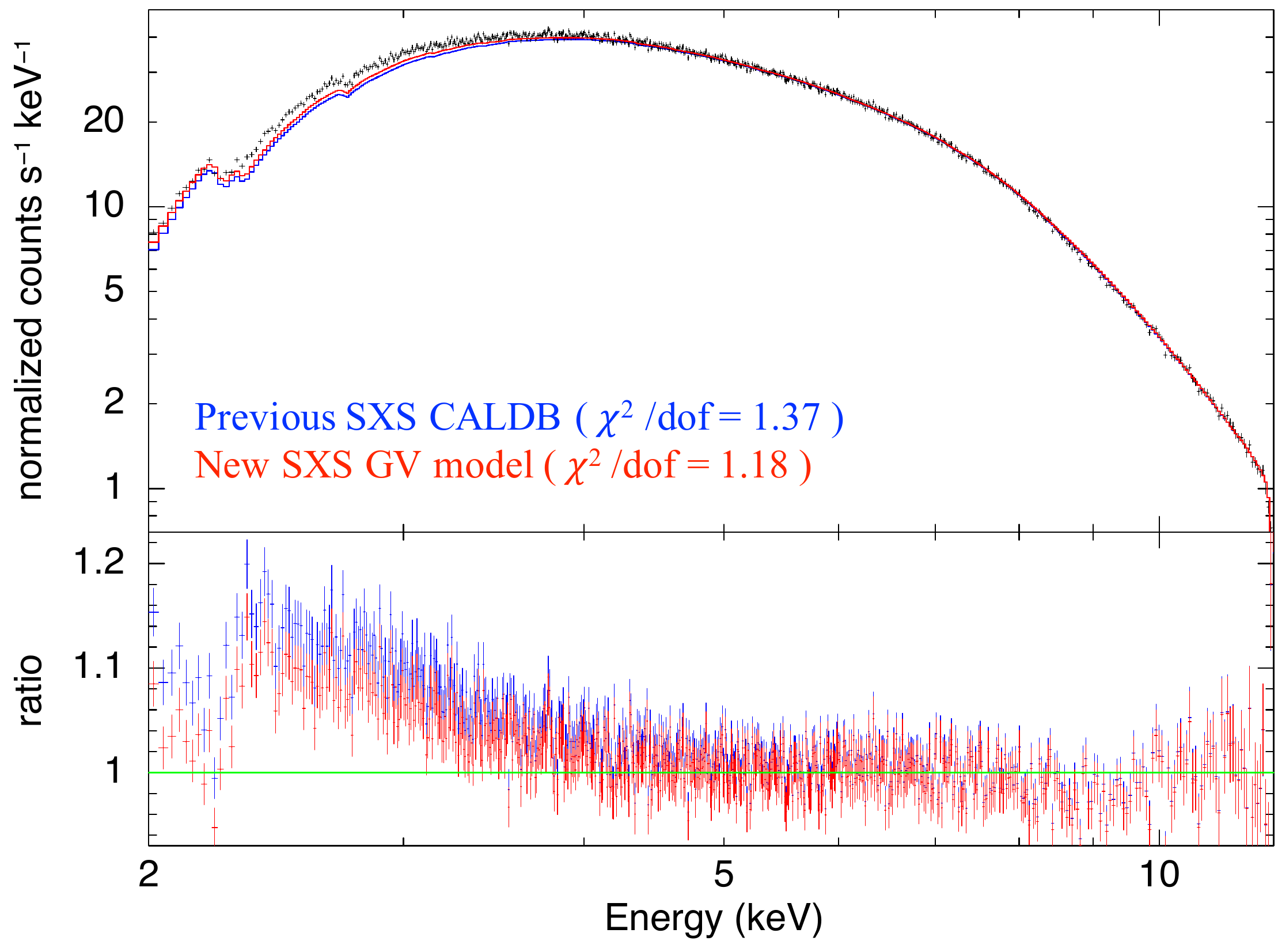}
  \caption[Comparison of simulated Crab spectra with two different ARFs and the observed
  data with the SXS]{Comparison of simulated Crab spectra with two different ARFs (blue
  and red) and the observed
  data with the SXS (black). The blue and red curves respectively show the models using
  the latest SXS CALDB and the one developed in this study, where the best Be thickness
  is 261~\um.}
  \label{fig:Crab}
 \end{center}
\end{figure}

Figure~\ref{fig:Crab} shows the comparison between the data and the model in the SXS
CALDB (blue) and the best one (261~\um) among our deviated thicknesses (red). We
confirmed that the best model better describes the Crab observation results. Further
improvement is hampered by other sources of calibration uncertainties such as Au M edge
features by the X-ray mirrors.

\section{SUMMARY AND CONCLUSION} \label{sec:summary}
We performed the X-ray transmission measurements of the two main components of the
\textit{Resolve} GV: the Be window and the stainless steel mesh. The Be window of the
\textit{Resolve} FM was measured in the 2.0--12.0~keV using the synchrotron facility at
KEK PF. The FS stainless steel mesh of the SXS was measured at six discrete energies
using the X-ray beamline at ISAS. The stainless steel mesh was measured for the first
time in this study.

We developed the X-ray transmission model of the GV. For the stainless steel mesh, the
wire cross section was changed from a square in the SXS model to a circle. The mesh
opening fraction was derived by fitting the data rather than being fixed to the design
value. For the Be window, the following factors were modeled: (1) the photoelectric
absorption and incoherent scattering of Be, (2) the photoelectric absorption of
impurities (Cr, Mn, Fe, Ni, and Cu), and (3) the coherent scattering of Be. The physical
interpretation of the transmission discontinuity caused by the Bragg diffraction was
presented and incorporated into the model. We also measured a spatial non-uniformity in
the Be window transmission. All these measurements and models satisfy the calibration
requirements.

In addition to constructing the transmission model of the \textit{Resolve} GV, we
revisited the models of the SXS GV based on the approach developed in this work. We
performed the reanalysis of the Crab spectrum observed with the SXS when the GV was
closed using the ground measurement data and revising the model parameters. We found
that the new model better explains the Crab observation results.

In this study, we performed an elaborated X-ray transmission modeling of the Be window
considering the beryllium's poly-crystal structure. We also showed utility of measuring
X-ray diffraction pattern in addition to the X-ray transmission. We expect that the
presented result will serve as a useful reference for all future X-ray spectrometers
with a Be window.

\clearpage
\appendix

\section{BDF energies}
\label{sec:appendix}
\setcounter{equation}{0} 
\renewcommand{\theequation}{\Alph{section}.\arabic{equation}}
\setcounter{figure}{0} 
\renewcommand{\thefigure}{\Alph{section}.\arabic{figure}}
\setcounter{table}{0} 
\renewcommand{\thetable}{\Alph{section}.\arabic{table}}

Table \ref{tab:all_BDF} summarizes all crystal planes ($hkl$) of Be below 12.0~keV,
their spacing $d_{hkl}$, multiplicity $M$, and the BDF energy $E_{\rm BDF}$ of the
($hkl$) plane. We also label whether each plane satisfies the systematic absence
condition for the Be hexagonal crystal that
\begin{equation}
-\frac{h}{3} + \frac{k}{3} + \frac{l}{2} = \frac{2n-1}{2},
\end{equation}
where $n$ is an integer.

\begin{table}[H]
\centering
\caption{Crystal information for crystal planes of Be below 12.0~keV}
\label{tab:all_BDF}
\scalebox{0.9}{
\begin{tabular}[t]{ccccc}\hline
$hkl$ & $d_{hkl}$ [\AA] & $E_{\rm BDF}$ [keV]  & $M$ & Systematic absence \\\hline
001	&	3.5842	& 1.7298 &	2 & absent \\
100	&	1.9791	& 3.1326	& 6 &  \\
002	&	1.7921	& 3.4596	& 2 &  \\
101	&	1.7325	& 3.5785	& 12 &  \\
102	&	1.3284	& 4.6671	& 12 &  \\
003	&	1.1947	& 5.1894	& 2 & absent \\
110	&	1.1427	& 5.4259	& 6 &  \\
111	&	1.0887	& 5.6949	& 12 & absent \\
103	&	1.0228	& 6.0616	& 12 &  \\
200	&	0.9896	& 6.2653	& 6 &  \\
112	&	0.9635	& 6.4350	& 12 &  \\
201	&	0.9539	& 6.4997	& 12 &  \\
004	&	0.8961	& 6.9191	& 2 &  \\
202	&	0.8663	& 7.1570	& 12 &  \\
113	&	0.8258	& 7.5080	& 12 & absent \\
104	&	0.8163	& 7.5953	& 12 &  \\
203	&	0.7621	& 8.1353	& 12 &  \\
210	&	0.748	& 8.2882	& 12 &  \\
211	&	0.7323	& 8.4668	& 24 &  \\
005	&	0.7168	& 8.6489	& 2 & absent \\
114	&	0.7051	& 8.7929	& 12 &  \\
212	&	0.6903	& 8.9812	& 24 &  \\
105	&	0.674	& 9.1988	& 12 &  \\
204	&	0.6642	& 9.3342	& 12 &  \\
300	&	0.6597	& 9.3979	& 6 &  \\
301	&	0.6488	& 9.5558	& 12 & absent \\
213	&	0.634	& 9.7787	& 24 &  \\
302	&	0.6191	& 10.0145	& 12 &  \\
115	&	0.6072	& 10.2100	& 12 & absent \\
006	&	0.5974	& 10.3787	& 2 &  \\
205	&	0.5805	& 10.6798	& 12 &  \\
303	&	0.5775	& 10.7355	& 12 & absent \\
214	&	0.5742	& 10.7967	& 24 &  \\
106	&	0.5719	& 10.8412	& 12 &  \\
220	&	0.5713	& 10.8518	& 6 &  \\
221	&	0.5642	& 10.9888	& 12 & absent \\
310	&	0.5489	& 11.2949	& 12 &  \\
222	&	0.5443	& 11.3899	& 12 &  \\
311	&	0.5426	& 11.4266	& 24 &  \\
304	&	0.5313	& 11.6703	& 12 &  \\
116	&	0.5294	& 11.7114	& 12 &  \\
312	&	0.5248	& 11.8128	& 24 & \\
215	&	0.5176	& 11.9791	& 24 &  \\\hline
\end{tabular}
}
\end{table}

\acknowledgments 
The authors acknowledge Megan E. Eckart (LLNL), Maurice A. Leutenegger (NASA GSFC), Yoh
Takei (JAXA), and anonymous reviewers for their valuable comments. This work was
performed under the approval of the Photon Factory Program Advisory Committee (Proposal
No. 2018G509) and Proposal Assessing Committee of HiSOR (Proposal No. 19BU007). We
acknowledge Shinjiro Hayakawa (Hiroshima University) for the upgrading the HiSOR BL-11
beamline and supporting our measurements. We appreciate Koichi Kitazono and Yuta
Fujimori (Tokyo Metropolitan University) for allowing us to use their X-ray
diffractometer and supporting our experiments. We used software, API, and databases
provided by the Cambridge Crystallographic Data Centre. T.M is financially supported by
JSPS Grant-in-Aid for JSPS Research Fellow (JP20J20809). A part of the results of this
manuscript was previously reported in SPIE proceedings \cite{Midooka20a}.


\bibliography{spiejour}   
\bibliographystyle{spiejour}   


\vspace{2ex}\noindent\textbf{First Author} is a PhD candidate at the University of
Tokyo. He received his BS degree in physics from Kyoto University in 2018, and his MS
degree in astronomy from the University of Tokyo in 2020. His current research interests
include X-ray and gravitational wave astronomy.

\vspace{1ex}
\noindent Biographies and photographs of the other authors are not available.

\listoffigures
\listoftables

\end{spacing}
\end{document}